\documentstyle[osa,aps,preprint]{revtex}
\begin{document}
\draft

\title{Mode-Locking Hysteresis in the Globally Coupled
Model of Charge-Density Waves}
\author{Afshin Montakhab, J. M. Carlson, and Jeremy Levy}
\address{Department of Physics, University of California,
Santa Barbara, CA 93106-9530}

\maketitle

\begin{abstract}
We study the response of a recently proposed
global coupling model of charge-density waves
to a joint ac+dc drive.  The model is the
standard Fukuyama-Lee-Rice model with an
additional global coupling term to account for
interaction with (uncondensed) quasi-particles.
We find that traditional mode-locking
is accompanied by hysteresis on mode-locked steps.
The experimentally observed asymmetry in the
size of hysteresis is reproduced and explained
by drawing attention to the importance of
the pinning force.   For large values of
global coupling, increased
rigidity leads to vanishing of
subharmonic steps consistent with experiments.

\end{abstract}

\pacs{PACS numbers: 72.15.Nj, 05.45.+b}

\narrowtext

Charge-density wave (CDW) conductors
have  been established
as ideal systems in which to  study
the dynamics of many
degrees of freedom  in the presence of
disorder \cite{cdwreview}.  CDWs display a variety
of nonlinear nonequilibrium phenomena,
including dynamical phase transitions \cite{dcf83}
and mode-locking \cite{mlreview}.
In the presence of an applied dc-field $E_{dc}$,
above a threshold $E_{th}$,
the CDW depins and carries an average current
$J_{cdw}$ which  varies continuously
with $E_{dc}$.
A similar behavior is observed
at the  mode-locking transitions,
where in the presence of  joint ac+dc drive
the ``narrow-band noise''
frequency, $\omega_{nbn}\propto J_{cdw}$, which
characterizes the CDW current oscillations, locks onto
the external frequency $\omega_{ex}$
with
$\omega_{nbn}/\omega_{ex}=p/q$,  $p$ and
$q$ integers.
The mode-locked state is the analog of the pinned state,
with the  CDW remaining locked at the same frequency
for a finite range of   $E_{dc}$.
This defines
a mode-locked step over which the average
CDW current  remains fixed.
Upon unlocking at  lower and higher  ``thresholds''
the current varies continuously with $E_{dc}$.

Over a  range of temperatures
the depinning  and mode-locking
transitions are
smooth and  can be analyzed in terms
of dynamical critical phenomena \cite{dcf83}.
However,
at lower temperatures, CDWs
also display discontinuous and sometimes
hysteretic transitions.
In the case of the depinning transition,
hysteresis, or \lq\lq switching," is observed
at a second, higher threshold field
$E_{th}^*$ \cite{adelman93}.
More recently,
Higgins {\it et al.} \cite{higgins93}
observed hysteresis on mode-locked steps. Most notably,
the size of the hysteresis  loops on opposite
edges of a given step
exhibit a
marked asymmetry: for certain
values of the ac-drive the right edge of
the step  is hysteretic while the left edge
displays a smooth mode-locking transition.
The observed hysteresis is
small in size and occurs in a
parameter regime in which the sample
is not switching in
the absence of ac-drive.
In a different set of experiments,
Levy {\it et al.} \cite{levychaos93}
pointed to the chaotic
behavior on the mode-locked steps in
switching samples, in which case no well-defined
mode-locking hysteresis loop can be identified.
Instead, a period-doubling
route to chaos with an associated
low-dimensional attractor was  observed
while at lower frequencies
high dimensional and noisy dynamics were
seen.

These hysteretic and chaotic phenomena
provide new challenges
to our theoretical understanding of these systems.
In particular,
motivated by these
experiments, we have studied the mode-locking
behavior of a new global coupling model \cite{globalcoupling}
of CDW dynamics which
has been seen previously to show some promise in explaining the
temperature dependent behavior of CDWs  by producing
hysteresis at the depinning transition
as well as delayed conduction
\cite{delay93}.
In this paper we focus our attention
on the one-dimensional version
of this model, which we find
produces some of  the experimentally observed
results and also allows us to make  further predictions;
however, the 1d model comes up short
on some important aspects of the experiments as well.

Many aspects of CDW dynamics including
the  phenomena
of mode-locking are well described by the Fukuyama-Lee-Rice
(FLR) model \cite{flr}. An important property of this model
is the so-called no passing rule \cite{nopass}
which requires that
the current   be a single-valued function of $E_{dc}$.
Hence the FLR model
cannot  describe the low-temperature
hysteresis exhibited at the depinning or
mode-locking transitions.  As
pointed out by Littlewood \cite{littlewood88}
and recently shown by Levy {\it et al.} \cite{globalcoupling},
the role of normal carriers
excited across the CDW gap ($\bigtriangleup_{CDW}$)
becomes increasingly
important at low temperatures.
This adds
a new global coupling term to the
FLR equations of motion,
altering the I-V characteristics
significantly.

The global coupling
model is obtained by  incorporating
current conservation within the FLR equation of
motion.
Ignoring the displacement current,
the conservation equation is readily
integrated in 1d \cite{globalcoupling}
leading to the following equation suitable
for numerical simulations,
\begin{eqnarray}
(\gamma_0+\gamma_1) \dot{\phi_i}=&&
k\bigtriangleup^2 \phi_i-\sin(2 \pi(\phi_i
+\beta_i))+E(t)\nonumber\\
&& +
{\gamma_1 \over N}
\sum^{N}_{i=1} \dot{\phi_i} .
\end{eqnarray}
Here $\phi_i$ represents the CDW phase at the $i$th
impurity site, $\gamma_0$ is the time constant for
CDW relaxation (we set $\gamma_0=1$),
and $\gamma_1$ is the normal
carrier resistance, assumed to have an exponential
temperature-dependence:
$\gamma_1 \propto \exp  (\bigtriangleup_{CDW}/k_BT)$.
The FLR pinning parameter is $k$,
so that the two limits
$2 \pi k \ll 1$ and $2 \pi k \gg 1$
correspond to the strong and weak pinning limits,
respectively.  The
$\beta_i \in [0,1)$
are random variables  due to disorder, and
$\bigtriangleup^2$ is the lattice Laplacian.
The average CDW
current is given by $\bar{J}_{cdw}=<\dot{\phi}_i>_{x,t}
=E_{dc}+\bar{F}_{pin}$ where the second term
is the time and space averaged pinning force
\begin{eqnarray}
\bar{F}_{pin}= -<\sin(2 \pi(\phi_i + \beta_i))>_{x,t}.
\end{eqnarray}
We simulate Eq.~(1) with
$E(t)=E_{dc} + E_{ac}\cos(\omega_{ex}t)$
using periodic boundary conditions in
one dimension.  Fifth-order Cash-Karp Runge-Kutta
\cite{recipes92} is used to integrate the
equations for $N$ varying from $10$ to
$1024$ with various values of $k$.

When $\gamma_1=0$  the standard
FLR model is recovered.  For non-zero $\gamma_1$ the
time constant for $\dot{\phi}_i$ is
adjusted and  the
global coupling term is relevant.  It was
shown \cite{globalcoupling} that the additional term gives
rise to hysteresis
in the regular ($E_{ac}=0$) I-V curve;
however, the size of hysteresis  can be very small
for  small $\gamma_1$, with the difference between the
pinning and depinning fields  scaling roughly as
$\gamma_1^2$ for $\gamma_1 \ll 1$ \cite{kurt}.
For the mode-locking transition, we define the size
of hysteresis $S_{hyst}$ to be the magnitude of
the difference between the locking and unlocking
fields, and examine variations in $S_{hyst}$ in
different regimes.

Figure 1 shows  mode-locking ($E_{ac} \neq 0$)
I-V curves for
different values of $\gamma_1$ in the
weak pinning regime ($k=1.0$).  For
$\gamma_1 = 0.0$  harmonic ($q=1$) and
many high order subharmonic ($q \neq 1$) steps are produced.
As $\gamma_1$  increases
hysteresis emerges asymmetrically on harmonic and some
subharmonic steps.  With increasing $\gamma_1$
subharmonic steps become smaller as  harmonic steps
grow wider.  The left edges of the steps become rounded
as the right edges become sharp and hysteretic.
No chaotic behavior is observed.
Note that within a hysteresis loop (e.g.
right edge of the
0/1 step in Fig. 1(b)) two distinct mode-locked states,
i.e. different $p/q$ ratios,
sometimes exist for identical set of parameters, one on the
upper branch and one on the lower branch.  This means
that nondegenerate mode-locked
attractors exist for this
model, where  the FLR model
only exhibits degenerate mode-locked
attractors displaying the same velocity
\cite{nopass}.
We emphasize that  in order to obtain
this much structure in the I-V curves it  was
necessary to go to relatively large values of $k$.
In comparison,
while
the strong pinning case of $k=0.01$  also displays
asymmetric hysteresis on mode-locked steps,
the I-V structure
was much simpler, displaying
harmonic and only $p/2$ subharmonic steps
even for large system sizes.

In analogy with the behavior  for the
depinning transition \cite{globalcoupling},
we expect the $S_{hyst}$ to increase
with  increasing $\gamma_1$. This is always observed
for strong pinning, and in general this
characterizes the basic trends.
However, in the weak pinning case, there
are sometimes  exceptions occurring in
cases where we observe subharmonic steps on the upper branch
of the hysteresis loops.
These additional  steps within a hysteresis loop
appear to increase
$S_{hyst}$ as is seen by comparing
Fig. 1(b) and 1(c). This feature disappears
for small $k$ or large $\gamma_1$.

In general, the disappearance of subharmonic steps
throughout the I-V curve with
increasing $\gamma_1$ is consistent with
experiments \cite{levychaos93}.
In the model internal deformations of
the CDW must be accompanied by the screening
currents of quasi-particles in order
to conserve charge locally.  With decreasing
temperature quasi-particle conductivity ($1/\gamma_1$)
decreases exponentially,  inhibiting
the ``backflow'' current. This freezes out
the internal CDW deformations and
results in a rigid CDW.  It is well known that
subharmonic mode-locking is a feature of
many degrees-of-freedom dynamics and that
this behavior disappears in single degree-of-freedom
models\cite{mlreview}.
Here, the disappearance is due
to low dimensional behavior in a high
dimensional model,  arising because  of the dominance
of the global coupling.

Figure 2 shows the ``phase diagram'' in the weak
pinning ($k=1.0$) regime for $\gamma_1=2.5$.
Figure 2 differs from the  standard
phase diagrams where the step width is plotted
in the
$E_{ac}$ vs.~$E_{dc}$  plane.
Here, in order to
illustrate the existence and variations of
hysteresis, we  plot the full I-V curves for different
values of $E_{ac}$.  For each mode-locked
state the width of the step oscillates as
$E_{ac}$ is varied.
Within the FLR model,
in the high velocity limit, the oscillations
in  harmonic mode-locked
steps  are described by
Bessel functions \cite{bessel}.
In Fig. 2 the dashed lines  connecting the  unlocking
transition points appear to
be at least qualitatively consistent
with the FLR case
though we have not investigated the functional form in detail.
The hysteresis size asymmetry is most pronounced
in the first half of the first oscillations
(low $E_{ac}$)
where large hysteresis is observed.   For
larger $E_{ac}$, as in the second oscillation
of the 1/1 step, hysteresis becomes much
smaller and symmetric on the opposite  edges of the steps.
The asymmetry
occurs even for high velocity mode-locked steps, e.g.
4/1 step (not shown here), at low $E_{ac}$ values.
Therefore, asymmetry appears to be a low $E_{ac}$
but not necessarily a low velocity
phenomenon.

We suggest that
an important variable governing
the size of hysteresis within a given phase diagram
is the pinning force at
the edges of the mode-locked step.
In Fig. 3 we show a schematic asymmetric
mode-locked step labeling important
quantities.
By drawing the $\bar{J}_{cdw} = E_{dc}$ line through
the step, one can visualize the magnitude  of
the average pining force,
$|\bar{F}_{pin}|$, as the height difference between the
two curves.
One might
naively expect that in the middle of the
step the pinning force would be zero so that at
the edges of the step the pinning force would
take on some extremum values as allowed by the value
of $E_{ac}$.  This is indeed the case for  large
$E_{ac}$; the second
oscillations, for example, exhibit
steps for which the
value of the pinning force at the
center of the step is close to zero, and the value of
$\bar{F}_{pin}^{edge}$ is equal in magnitude,
with opposite sign, at the
two edges.  However, within the
first oscillation, the center of the steps
does not occur at $E_{dc}=\bar{J}_{cdw}^{p/q}$
i.e. where $\bar{F}_{pin}=0$.  The centers of these
steps lie at higher values of $E_{dc}$
(nonzero pinning force) and this ultimately gives rise to
the hysteresis asymmetry of the steps.  The
center of these steps occurs at higher values
of $E_{dc}$ since for small $E_{ac}$
the pinning force cannot change appreciably from
its value when $E_{ac}=0$.  To increase
$\bar{J}_{cdw}$ to the mode-locked
value, $\bar{J}_{cdw}^{p/q}$,
one must go to higher $E_{dc}$. Therefore,
for low $E_{ac}$ the steps are ``shifted'' to
the right as is apparent from Fig. 2.

The above argument suggests that the 0/1
step should always be very nearly symmetric,
since for the 0/1
step the pinning force is  zero
at $E_{dc}=0$, which is always close to
the center of the step \cite{comment}.
Higgins
{\it et al.} \cite{higgins93}
did not look at the left edge of the
0/1 step so this remains to be verified.
However, they did
discover that for steps other than
0/1, hysteresis occurred for
those values of $E_{ac}$ where the steps
were the widest, i.e.
middle of the first oscillation.
This is consistent with Fig. 2
where the largest
hysteresis occurs in the middle of
the first oscillation.
More generally, the wider steps tend to
have a larger pinning force at the edges,
and the larger the pinning force the larger
the size of hysteresis.

The observed qualitative trends
in the relationship between
the size of   the hysteresis
(e.g.~$S_{hyst}=E_{T_2}-E_{T_3} $
for the right edge hysteresis  in Fig.~3)
and the magnitude of the  pinning force (Eq.~(2))
suggests that it may be worth
searching for some quantitative scaling  relationship.
Restricting our attention to a particular
phase diagram in the strong pinning regime
($k=0.01$ and $\gamma_1=1.0$, and $w_{ex}=2\pi$),
we have attempted to find such a
relationship with limited success.
In particular,
a scaling fit
between $S_{hyst}$ and  $\bar{F}_{pin}^{edge}$
appears to hold well for the symmetric steps.
However, an extrapolation of this scaling
breaks down in the low $E_{ac}$ limit where
the asymmetry is most pronounced.
In that case on the right edges of the steps,
$S_{hyst}$ is
even larger than what the extrapolation predicts,
while $S_{hyst}$ is smaller
(often zero) on the left edges.
This  result is indicative of
a crossover occurring between the high
and low $E_{ac}$ limits, which can be
accounted for in a crude manner using the
pinning force at the center of the step
$\bar{F}_{pin}^{center}$. Note that
$\bar{F}_{pin}^{center}$
is  negative at low $E_{ac}$  increasing to
zero at high $E_{ac}$.
We construct an adjusted pinning force
$\bar{F}_{pin}^{adj}
=\bar{F}_{pin}^{edge}+
\alpha \times \bar{F}_{pin}^{center}$.
Note that
$\bar{F}_{pin}^{adj}=
\bar{F}_{pin}^{edge}$
for symmetric steps.
However, for asymmetric steps with $\alpha > 0$,
$|\bar{F}_{pin}^{adj}|$ is increased on the right edge
and decreased on the left edge
as compared to
$\bar{F}_{pin}^{edge}$, yielding a
more satisfactory fit
to the data.  The result is shown in Fig. 4 for the
2/1 and 1/1 steps where
$S_{hyst} \sim |\bar{F}_{pin}^{adj}|^{\lambda}$
roughly describes the data. Here, instead
of fitting the data to
find the best values of $\alpha$ and $\lambda$ we
have plotted the result for $\alpha=5.0$.

Finally, regarding the  discrepancies
between our results
and those obtained experimentally,
we note that in all the simulations,
for each given ``sample'', the
largest hysteresis typically occurs  when $E_{ac}=0$;
in other words,  mode-locking hysteresis
occurred only in ``switching samples.''
Furthermore, no chaos was observed for a
wide range of parameter values.  This
is troublesome in
the light of the experiments which show hysteresis
in non-switching samples,
and chaos at the step edges of  switching samples.
One possible resolution of this
may come from further study of the depinning transition.
Real CDWs exhibit two transitions, only one of
which becomes hysteretic at
low temperatures \cite{adelman93}.
The 1d model studied here exhibits  one
transition
where  a small hysteresis
grows to a large one as a function
of $\gamma_1$.  However, there
is evidence \cite{kvale} that a 2d generalization
of the model  exhibits depinning
transitions  which  more closely
resemble the experiments.
In that case
a new ``creep'' motion
arises  between the first and second
transitions, and perhaps this qualitatively different
behavior will
bare out
these shortcomings.

In summary, we have studied the
response  of the global
coupling model  to a joint
ac+dc field.
Unlike the FLR model, this model
exhibits  hysteresis and nondegenerate
mode-locked attractors.   The model produces hysteresis
on mode-locked steps exhibiting size asymmetry
in the low ac-field regime.
The pinning force and the step offset
are the primary variables which control mode-locking
hysteresis within a given phase diagram.  Increasing
rigidity of the CDW at large values of
global coupling leads to low dimensional
behavior and thus vanishing of subharmonic
steps.
While the model exhibits no chaos
in the regime studied,
this low dimensional behavior may
have some bearing on the
low-dimensional chaotic
attractors observed experimentally \cite{levychaos93}.

It is a pleasure to acknowledge useful
conversations with G. Fiddich,
M. Kvale, J. McCarten,
C. Myers, and M.S. Sherwin.  A.M. acknowledges
J. Cleveland for help with graphics.
Computations were performed in part at
the San Diego Supercomputer Center.
This work is supported by fellowships from
the Alfred P. Sloan Foundation, the David
and Lucile Packard Foundation, and NSF grants
DMR-9212396 and DMR-9214995.

\begin{figure}
\caption{Mode-locking I-V curves for different values of
the global coupling parameter,
$\gamma_1$. Note that (a) is the
FLR limit.  The curves are  shifted
vertically for clarity.
Some steps are labeled by their p/q
values in (a).  The emergence of asymmetric
hysteresis with increasing global coupling
is evident.\label{figure1}}
\end{figure}

\begin{figure}
\caption{Mode-locking ``phase diagram''; I-V
curves for different values of $E_{ac}$. Here $k=1.0,
N=256, \gamma_1=2.5$ and $\omega_{ex}=3.7$.
Curves are shifted vertically for
clarity. The lowest curve is the regular
I-V curve.
Dashed lines connect the
unlocking transition points for each step.
Hysteresis varies in size for each step and is
largest in the middle of the first
oscillation. Note that the steps  are  offset
to the right for low values of $E_{ac}$ and that the
offset decreases as $E_{ac}$ increases.\label{figure2}}
\end{figure}

\begin{figure}
\caption{Geometric representation of $\bar{J}_{cdw}=
E_{dc} + \bar{F}_{pin}$ on an asymmetric mode-locked step.
The step is drawn schematically to label
important quantities and illustrate the  nature of the
asymmetry.\label{figure3}}
\end{figure}

\begin{figure}
\caption{Hysteresis size vs.~adjusted pinning
force for 1/1 and 2/1 steps in the
strong pinning regime. $S_{hyst} \sim |\bar{F}^{edge}_{pin}+
\alpha \times \bar{F}^{center}_{pin}|^{\lambda}$ with
$\alpha=5.0$ (see text).
\label{figure4}}
\end{figure}


\begin{references}
\bibitem{comment} While simulations reveal that
the 0/1 step is nearly symmetric, this symmetry is
not exact. Note that Eq.~(1) is not symmetric
under $\phi_i \rightarrow - \phi_i$ and
$E(t) \rightarrow -E(t)$ since the
pinning term
breaks the symmetry.
\end{references}
\end{document}